\documentclass[a4paper,11pt]{article}
\pdfoutput=1
\usepackage{graphicx,rotating,hyperref,slashed,amsmath,xcolor,amssymb,amsfonts,colortbl,cite, subfigure,float,soul}
\makeatletter 
\usepackage{wrapfig}
 
\usepackage{graphics} 
\usepackage{graphicx} 
\graphicspath{{plots/}}
 
\hypersetup{colorlinks,bookmarksopen,bookmarksnumbered,
linkcolor=blus,pdfstartview=FitH,urlcolor=rossos,citecolor=verde}
\numberwithin{equation}{section}

\hypersetup{%
    ,urlcolor=rossos
    ,citecolor=rossos
    ,linkcolor=rossos
    }

\AtBeginDocument{
  \hypersetup{
    urlcolor=verdes,
    citecolor=verdes,
    linkcolor=verdes,
  }%
}

\allowdisplaybreaks

\def\lsim{\mathrel{\rlap{\lower3pt\hbox{\hskip0pt$\sim$}}
   \raise1pt\hbox{$<$}}}         
\def\gsim{\mathrel{\rlap{\lower4pt\hbox{\hskip1pt$\sim$}}
   \raise1pt\hbox{$>$}}}         

 \newcommand{\sfootnote}[1]{} 
\definecolor{bluc}{cmyk}{1,1,0,0.1}
\definecolor{rossoCP3}{cmyk}{0,.88,.77,.40}
\definecolor{rosso}{cmyk}{0,1,1,0.4}
\definecolor{rossos}{cmyk}{0,1,1,0.55}
\definecolor{rossoc}{cmyk}{0,1,1,0.2}
\definecolor{verdes}{cmyk}{0.92,0,0.59,0.4}

\hypersetup{colorlinks, bookmarksopen, bookmarksnumbered,
citecolor=verdes, linkcolor=bluc, pdfstartview=FitH, urlcolor=rossos}

\newcommand{\mio}[1]{}

\definecolor{Gray}{gray}{0.95}

\usepackage{multicol}
\usepackage{color}
\definecolor{rosso}{cmyk}{0,1,1,0.4}
\definecolor{rossos}{cmyk}{0,1,1,0.55}
\definecolor{rossoc}{cmyk}{0,1,1,0.2}
\definecolor{blu}{cmyk}{1,1,0,0.3}
\definecolor{blus}{cmyk}{1,1,0,0.6}
\definecolor{bluc}{cmyk}{1,1,0,0.1}
\definecolor{verde}{cmyk}{0.92,0,0.59,0.25}
\definecolor{verdec}{cmyk}{0.92,0,0.59,0.15}
\definecolor{verdes}{cmyk}{0.92,0,0.59,0.4}

\def\circa#1{\,\raise.3ex\hbox{$#1$\kern-.75em\lower1ex\hbox{$\sim$}}\,}

\newcommand{\beq}{\begin{equation}}
\newcommand{\eeq}{\end{equation}}

\newcommand{\bea}{\begin{eqnarray}}
\newcommand{\eea}{\end{eqnarray}}
\newcommand{\be}{\begin{equation}}
\newcommand{\ee}{\end{equation}}
\newfam\rsfsfam
\def\mathscr#1{{\fam\rsfsfam\relax#1}}

\def\circa#1{\,\raise.3ex\hbox{$#1$\kern-.75em\lower1ex\hbox{$\sim$}}\,}
\makeatletter

\def\hhref#1{\href{http://arxiv.org/abs/#1}{arXiv:#1}} 

\newcommand{\doi}[1]{\href{http://dx.doi.org/#1}{[doi]}}

\def\hhref#1{\href{http://arxiv.org/abs/#1}{arXiv:#1}} 
 
\def\art{\@ifnextchar[{\eart}{\oart}}
\def\eart[#1]#2#3#4#5#6{{\rm #2}, {\em #3 \bf #4} {\rm (#6) #5} ({\em #1})}

\def\article{\@ifnextchar[{\earticle}{\oarticle}}
\def\oarticle#1#2#3#4#5#6{{\rm #1}, {\em ``#6''}, {\rm #2 #3 (#5) #4}}
\def\earticle[#1]#2#3#4#5#6#7{{\rm #2}, {\em ``#7''}, {\rm #3 #4 (#6) #5}  [\hhref{#1}]}
\def\hepart[#1]#2{{\rm #2, \em#1}}
\def\heparticle[#1]#2#3{#2, {\em ``#3''} [\hhref{#1}]}

\newcounter{alphaequation}[equation]
\def\thealphaequation{\theequation\hbox to
0.6em{\hfil\alph{alphaequation}\hfil}}
\def\eqnsystem#1{
\def\@eqnnum{{\rm (\thealphaequation)}}
\def\@@eqncr{\let\@tempa\relax \ifcase\@eqcnt \def\@tempa{& & &} \or
  \def\@tempa{& &}\or \def\@tempa{&}\fi\@tempa
  \if@eqnsw\@eqnnum\refstepcounter{alphaequation}\fi
\global\@eqnswtrue\global\@eqcnt=0\cr}
\refstepcounter{equation} \let\@currentlabel\theequation \def\@tempb{#1}
\ifx\@tempb\empty\else\label{#1}\fi
\refstepcounter{alphaequation}
\let\@currentlabel\thealphaequation
\global\@eqnswtrue\global\@eqcnt=0 \tabskip\@centering\let\\=\@eqncr
$$\halign to \displaywidth\bgroup \@eqnsel\hskip\@centering
$\displaystyle\tabskip\z@{##}$&\global\@eqcnt\@ne
\hskip2\arraycolsep\hfil${##}$\hfil& \global\@eqcnt\tw@\hskip2\arraycolsep
$\displaystyle\tabskip\z@{##}$\hfil
\tabskip\@centering&\llap{##}\tabskip\z@\cr}
\def\endeqnsystem{\@@eqncr\egroup$$\global\@ignoretrue} \makeatother

\definecolor{fiorentina}{rgb}{.5,0,.5}

\begin{document}

\vspace{1cm}
\begin{center}

{\fontsize{20}{28}\selectfont  \sffamily \bfseries {Inflationary magnetogenesis beyond slow-roll\\
and its induced gravitational waves}}

\end{center}

\vspace{-0.1cm}
\renewcommand{\thefootnote}{\fnsymbol{footnote}}

\hskip0.2cm
\begin{center} 
{\fontsize{13}{30}
 Bill Atkins$^{a\!}$,  Debika Chowdhury$^{b\!}$, Alisha Marriott-Best$^{a\!}$ , Gianmassimo Tasinato$^{a,c\!}$
} 
\end{center}


\begin{center}

\vskip 6pt
\textsl{$^a$ Physics Department, Swansea University, SA2 8PP, UK}
\\
\textsl{$^b$ Indian Institute of Astrophysics, II Block, Koramangala, Bengaluru 560034, India}
\\
\textsl{$^{c}$ Dipartimento di Fisica e Astronomia, Universit\`a di Bologna,\\
 INFN, Sezione di Bologna,  viale B. Pichat 6/2, 40127 Bologna,   Italy}
\vskip 4pt

\end{center}
\hskip0.1cm

\begin{abstract}
\noindent
The origin of magnetic fields observed on both astrophysical and cosmological scales is a compelling problem that has the potential to shed light on the early Universe. 
    We analytically investigate inflationary magnetogenesis in scenarios where a brief departure from slow-roll inflation -- akin to mechanisms proposed for primordial black hole formation -- leads to enhanced magnetic field generation with a growing power spectrum.
    Focusing on the Ratra model, we derive an analytic bound on the growth of the magnetic field power spectrum in this context, showing that the spectral index can reach \( d \ln {\cal P}_B / d \ln k = 4.75 \) during the growth phase.
This growth enables amplification from CMB-safe large-scale amplitudes to values of astrophysical relevance. We further compute the stochastic gravitational wave background sourced by the resulting magnetic fields, incorporating their rich spectral features. Under suitable conditions, the induced signal exhibits a characteristic frequency dependence and amplitude within reach of future gravitational wave observatories, providing a distinctive signature of this mechanism and a specific class of templates for upcoming gravitational wave searches.
\end{abstract}

\renewcommand{\thefootnote}{\arabic{footnote}}
\setcounter{footnote}{0}

\section{Introduction}
\label{sec_intro}

Understanding the origin of magnetic fields observed on large astrophysical and cosmological scales is a fascinating problem that may provide valuable insights into the early evolution of the Universe. Over the past few decades, significant efforts have been devoted to this question — see, for instance, the reviews in \cite{Kronberg:1993vk,Grasso:2000wj,Widrow:2002ud,Kandus:2010nw,Widrow:2011hs,Durrer:2013pga,Subramanian:2015lua}.
An intriguing possibility is that cosmic inflation generates the seeds for the subsequent evolution of cosmic magnetic fields: references \cite{Turner:1987bw,Ratra:1991bn,Martin:2007ue,Demozzi:2009fu,Kanno:2009ei,Bamba:2006ga,Barnaby:2012tk,Ferreira:2013sqa,Ferreira:2014hma} include original and influential works on primordial magnetogenesis. However, as discussed in these works, scenarios of inflationary magnetogenesis face stringent experimental and theoretical constraints that should be carefully addressed.
We adopt the scenario of Ratra \cite{Ratra:1991bn} 
and consider a  coupling of the inflaton field to the electromagnetic  field during inflation, 
so as to break the conformal invariance of the Maxwell action, and to allow for magnetic field production. 
We  link magnetogenesis to the physics of primordial black holes (see \cite{Ozsoy:2023ryl} for a
review), by  assuming that a brief phase of violation
of slow-roll conditions occurs during inflation, causing the inflaton velocity to change abruptly during a short period of time. Such non-slow-roll phase -- used in the context of
primordial black hole physics to enhance the spectrum of curvature fluctuations -- drastically affects the  time dependence of the coupling of the inflaton with Maxwell gauge fields, and influences the production of primordial magnetic fields at small scales only, with
interesting phenomenological ramifications. This scenario has the potential to selectively amplify a primordial magnetic field at certain small scales.  This topic has been recently investigated in \cite{Tripathy:2021sfb,Tripathy:2022iev}, though without addressing the specific questions we consider here.

Assuming an initial nearly scale-invariant magnetic field spectrum on large scales, with an amplitude low enough to satisfy stringent  constraints from  cosmic microwave background (CMB) observations \cite{Planck:2015zrl}, we show that a non-slow-roll epoch can rapidly amplify the magnetic field to levels compatible with those observed in astrophysical contexts.
We address two questions:
\begin{enumerate}
\item 
Within this approach,  what is the maximal possible slope of the magnetic field spectrum as it grows from large to small scales? This question is important  because the magnetic field amplitude must increase by several orders of magnitude from  tiny values at the largest  scales -- where it is constrained by CMB observations -- to smaller cosmological or astrophysical scales, where observations suggest much stronger fields.
An analog  problem has been studied in \cite{Byrnes:2018txb} in the context of adiabatic curvature perturbations \(\zeta\) relevant for the formation of primordial black holes. There, the amplification of \(\zeta\) from large to small scales was shown to be bounded by  
\(
\frac{d \ln {\cal P}_\zeta}{d \ln k} = 4,
\)
up to subdominant logarithmic corrections \cite{Carrilho:2019oqg,Ozsoy:2019lyy}.\footnote{Stronger growth may be possible if multiple successive phases of non-slow-roll evolution are allowed (see \cite{Tasinato:2020vdk}) or if the initial vacuum deviates from the Bunch–Davies form (see \cite{Cielo:2024poz}).}
In Section~\ref{sec_maxs}, we address the analogous question for the magnetic field spectrum. We develop a fully analytical method, building on \cite{Tasinato:2020vdk,Tasinato:2023ukp}, and find that the growth rate of the magnetic field spectrum \({\cal P}_B\) can exceed that of the curvature spectrum. In fact, under our assumptions, the maximal slope is 
\(
\frac{d \ln {\cal P}_B}{d \ln k} = 4.75,
\)
and the resulting magnetic spectrum has a rich
 profile as a function of the momentum scale. 
 We also comment how our scenario addresses
 strong coupling and backreaction problems
 of magnetogenesis. 
\item 
Given that our non-slow-roll mechanism leads to a rapid  amplification of the magnetic field spectrum at small scales,  does this process also generate vector-induced gravitational waves (GWs) at second order in magnetic field fluctuations?
The generation of GWs induced by adiabatic scalar fluctuations has a well-established history, 
 see
e.g.  
\cite{Matarrese:1992rp,Ananda:2006af,Baumann:2007zm,Saito:2009jt,Bugaev:2009zh,Espinosa:2018eve,Kohri:2018awv}: it  has recently gained renewed interest as a  probe of primordial black hole scenarios (see, e.g., \cite{Domenech:2021ztg} for a  review). In contrast, less attention has been given to GW production from non-adiabatic sources (see, e.g., \cite{Domenech:2021and, Passaglia:2021jla, Domenech:2023jve,Ozsoy:2023gnl,Marriott-Best:2025sez}), although there is existing literature on GWs arising from magnetogenesis scenarios (see, e.g., \cite{
Durrer:1999bk, Caprini:2001nb,Mack:2001gc,  Pogosian:2001np, Caprini:2003vc, Shaw:2009nf, Saga:2018ont,Bhaumik:2025kuj, Maiti:2025cbi}).
A common assumption in previous studies of magnetically-induced GWs is that the magnetic field spectrum follows a simple power-law behavior. However, as outlined  above, in our scenario the magnetic spectrum exhibits a much richer structure. This requires an extension of the standard formalism of magnetically induced GW to properly account for the non-trivial spectral features of our setup.
In Section \ref{sec_gwa}, we present such an extension and compute the resulting GW spectrum. We find that, under suitable conditions, the amplitude of the induced GWs can be large enough to be potentially detectable by future GW observatories. The resulting GW spectrum exhibits a distinctive frequency dependence, which could serve as a characteristic signature of our scenario, helping to distinguish it from other early universe sources of GWs.
\end{enumerate}
We conclude in section \ref{sec_out}, which is 
followed by the technical Appendix \ref{app_est}.

\section{The maximal slope of the magnetic field spectrum}
\label{sec_maxs}
We consider an Einstein-Maxwell Lagrangian for the electromagnetic field
in an expanding universe, 
\bea
\label{eq_emaxI}
S&=&\,\int d^4 x \,\sqrt{-g}\,
\left[\frac{R}{2}-
\frac{ I^2(\tau)}{4} \,F_{\mu\nu} F^{\mu\nu} \right].
\eea
 We set $M_{\rm Pl}=1$,  we use a mostly plus
metric signature $ds^2\,=\,a^2(\tau)\left[ - d \tau^2+d {\bf x}^2\right]$. We define $F_{\mu\nu} =\partial_\mu A_\nu-\partial_\nu A_\mu$ in terms of the vector potential.
The overall coupling  $I(\tau)$ depends
on  time and serves to break
the conformal symmetry of Maxwell action. $I(\tau)$  usually   follows a power law profile in terms of the scale
factor $a(\tau)$ during inflation.   Such behaviour is motivated by  directly coupling a function of the inflation field $\phi(\tau)$ with the Maxwell Lagrangian.  In a standard setup,  the scalar profiles change very slowly as functions of time, and a simple power-law behaviour for $I(\tau)$ as function of time 
can be easily be obtained. In this work we relax this assumption, motivating and analytically investigating scenarios where   $I(\tau)$
changes rapidly during inflation.

\smallskip

We decompose the Maxwell part of action \eqref{eq_emaxI} in terms of the spatial components $A_i$ of the vector potential (after integrating out the auxiliary time-like component $A_0$)
\bea
\label{eq_maxI}
S&=&
\frac12\,\int d \tau d^3 x \, I^2(\tau) \,
\left( 
A_i'^2-\frac12 \left( \partial_i A_j-\partial_j A_i \right)^2
\right).
\eea
The magnetic field components
scale with time as $B_i(\tau, {\bf x})\propto 1/a(\tau)$ \cite{Subramanian:2015lua}.  
 Hence after inflation we can write the
equality 
\be B_i(\tau, {\bf x})= a_R/a(\tau)\,B_i(\tau_R, {\bf x}),
\ee
where $B_i(\tau_R, {\bf x})\,=\,\epsilon_{ijk} \partial_j A_k(\tau_R, {\bf x})$ is the value of the magnetic field
right at the end of inflation.

The electro-magnetic potential is decomposed in Fourier space as
\be
\label{eq_fou}
A_i(\tau, {\bf x})\,=\,\sum_\lambda\,\int \frac{
d^3 {\bf k} }{(2 \pi)^{3/2}}\,e^{i
{\bf k}\cdot {\bf x}}\, {\bf e}_i^{(\lambda)}(\hat k)\,A_{k}(\tau),
\ee
with ${\bf e}_i^{(\lambda)}$ being two  polarisation vectors orthogonal to the momentum ${\bf k}$.  
  The amplitude  $P_{B}(k)$ of the 2-point
correlator for the magnetic field
at the end of inflation is defined
in terms of the Fourier transform of the magnetic field correlation function in Fourier space as follows:
\be
\label{eq_mfa}
\langle B_i(\tau_R, {\bf k}) B_j(\tau_R, {\bf q})\rangle_{{\bf k} ={\bf q}}'
\,=\,
\pi_{ij}\,P_{B}(k),
\ee
with 
\be
\pi_{ij}=\delta_{ij}-\hat k_i \hat k_j,
\ee
and the symbol $\langle\dots \rangle'_{\bf k=\bf q}$ indicates two-point correlators
omitting the momentum-conserving $\delta-$function. 
  From now on, vectors with a hat indicate
unit vectors, as $\hat k = {\bf k}/k$ with
$k=\sqrt{{\bf k}\cdot {\bf k}}$. 
To proceed, it is convenient to rescale $A_{k}$ by means of the conformal function $I$:
\be
{\cal A}_k(\tau)\,=\,I(\tau)\,A_k(\tau).
\ee
Its corresponding equation  of motion reads
\be
\label{eq_eom}
{\cal A}_k''+\left( k^2-\frac{I''}{I} \right)\,{\cal A}_k\,=\,0.
\ee
 Standard quantisation rules can be implemented (see e.g. \cite{subramanian2010, Tripathy:2021sfb,Martin:2007ue, Ferreira:2013sqa})
 leading to the following 
expressions for the
 magnetic
field 
energy density at the end of inflation -- we call it the $B$ spectrum: 
\bea
\label{eq_gmf}
{\cal P}_{B}(k)&=&
\frac{d \rho_{B}}{d \ln k} \,=\,
\frac{k^5}{2 \pi^2\,a^4_R} \left| {\cal A}_k\right|^2.
\eea
Comparing with eq \eqref{eq_mfa}, since 
$P_B\,=\,k^2 \left| {\cal A}_k\right|^2/I^2(\tau_R)$ at the end of inflation,
we find the following 
relation 
\be
\label{def_ampbs}
{\cal P}_B(k) \,=\, \frac{k^3\,I^2(\tau_R)\,P_B(k)}{2\,a_R^4\, \pi^2}.
\ee
As a warm-up
 example, we begin by selecting  $I(\tau)=a^2(\tau)$, and impose Bunch-Davies
initial conditions at small scales $|k \tau|\to \infty$. During inflation, we assume
de Sitter expansion $a(\tau)=-1/(H_I \tau)$, with $H_I$ being the constant Hubble parameter.
Equation \eqref{eq_eom} is easily solved analytically:
\be
{\cal A}_k(\tau)\,=\,\frac{3\,a^2(\tau)\,H_I^2}{\sqrt{2}\,
}\,
\frac{
e^{-i k \tau}
}{k^{5/2}} \left(1+ i k \tau-\frac{k^2 \tau^2}{3} \right).
\ee
At time  $\tau=\tau_R$, when inflation ends, the magnetic spectrum is 
\bea
\label{eq_pbst}
{\cal P}_{B}&=&\frac{9 H_I^4}{4 \pi^2} \left(1+ \frac{k^2 \tau_R^2}{3}+\frac{k^4 \tau_R^4}{9 } \right).
\eea
Working in the limit $|k \tau_R|\ll1$, this quantity becomes scale invariant. It
 is proportional to the fourth power of the inflationary Hubble parameter $H_I$,
 and can then be made very small at large scales (so as to satisfy
 existing stringent CMB bounds \cite{Planck:2015zrl}).

\medskip

Starting from the results of eq \eqref{eq_pbst},
 it is interesting to enquire whether there are mechanisms that would be able to amplify the
 magnetic spectrum at small scales, so as to possibly relate the observed magnetic fields at astrophysical or cosmological scales with primordial magnetogenesis. 

 \smallskip
 
 To do so, 
as anticipated in section \ref{sec_intro}, we  consider scenarios in which the function \( I(\tau) \) temporarily deviates from a power-law dependence on conformal time \( \tau \) during inflation. Such deviations can be motivated by inflationary models that include a brief phase of non-slow-roll evolution, which are particularly relevant in setups that produce primordial black holes. An appropriate
choice of the epoch 
where non-slow-roll occurs during inflation can enhance the magnetic field
amplitude at a convenient astrophysical or cosmological scale\footnote{Additionally,
  changes in the slope of \( I(\tau) \), as the ones we consider, can help in building magnetogenesis scenarios \cite{Ferreira:2013sqa} which avoid well-known strong coupling problems 
\cite{Demozzi:2009fu}. We elaborate further in section \ref{sec_strcou}. }.
Hence
in what follows we will dub this possibility {\it non-slow-roll magnetogenesis}.

 \smallskip

 We model \( I(\tau) \) as
\be
\label{eq_anin}
I(\tau)\,=\,a^2(\tau)\,\sqrt{\omega(\tau)},
\ee
with $\omega(\tau_R)=1$ when inflation ends. If we 
choose $\omega(\tau)=1$ throughout all the inflationary
evolution, we recover a scale invariant magnetic field spectrum
as outlined above\footnote{
We can consider other powers of the scale factor in Eq.~\eqref{eq_anin}, and write $I(\tau)\,=\,a^n(\tau)\,\sqrt{\omega(\tau)}$.
The analysis that follows can be readily extended to this generalised form.
}.
More generally,
extending the analysis of \cite{Tasinato:2020vdk} from
the scalar to the vector sector,
we parameterise
the  time-dependent function $\omega(\tau)$ 
  as
\bea
\label{ans_ome}
\omega(\tau)&=&
\begin{cases}
\omega_0 \,\,{\rm (a\,\, constant)}  & \text{for } \tau<\tau_1, \\ 
{\rm continuous\,\, but \,\,drastically\,\, changing}
&\text{for }  \tau_1<\tau<\tau_2,\\ 
 1& \text{for } \tau_2<\tau_R, 
\end{cases}
\eea
The two instants $\tau_1$ and $\tau_2$ during inflation are nearby, hence we assume
 $(\tau_1-\tau_2)/\tau_1\ll1$. Consequently, during the short
 time interval $ \tau_1<\tau<\tau_2$,  the
 conformal function $I(\tau)$ can experience strong departures from a power-law profile,
 as dictated by eq \eqref{eq_anin}. Correspondingly, a new characteristic scale 
 \be\label{eq_charsc}
 k_\star\,=\,-1/\tau_1\,=\,a(\tau_1) H_I
 \ee
 is  expected
 to play a relevant role in our discussion -- 
 such scale being  associated to a comoving momentum of modes 
 leaving the horizon at the onset of the non-slow-roll epoch during inflation. 
 We confirm this expectation in the following discussion.
 
 \subsection{Analytical determination
 of the magnetic mode function}

We now aim to solve equation~\eqref{eq_eom} with a general, slow-roll-violating function \( I(\tau) \) as given by equation~\eqref{eq_anin}.   
  In order to find analytic solutions of  eq
 \eqref{eq_eom} in the 
interval $ \tau_1<\tau<\tau_2$, we 
 proceed
as \cite{Tasinato:2020vdk}. (See also
\cite{Karam:2022nym,Franciolini:2022pav} for  different approaches to determine
analytic solutions for similar setup.)  We choose an Ansatz $\mathcal{A}_k(\tau)$
expanded as
\be
\label{eq_ans2}
{\cal A}_k(\tau)\,=\,\frac{3\,a^2(\tau)\,\sqrt{\omega(\tau)}
\,H_I^2}{\sqrt{2}}\,
\frac{
e^{-i k \tau}
}{k^{5/2}} \Big[1+ i k \tau - \frac{k^2 \tau^2}{3} + (i k \tau_0)^2 G_{(2)}(\tau) + (ik\tau_0)^3 G_{(3)}(\tau) + \dotsm \Big],
\ee
 with $\tau_0$ being a pivot quantity which will not appear in the final results, while $G_{(n)}(\tau)$
is a set of functions to be determined. 
Plugging this Ansatz into eq \eqref{eq_eom}, we find  a system
of coupled ordinary differential equations, which we aim to solve order by order in
powers of $k$:
\bea
\partial_\tau \left( \frac{\tau_0^2\,\omega}{\tau^4} \,G_{(2)}'\right)&=&\frac{\omega'}{3 \tau^3},
\\
\partial_\tau
 \left[
 \frac{\omega}{\tau^4} \left(\tau_0  \,G_{(3)}' - G_{(2)}
 -\frac{\tau^2}{3 \tau_0^2}
 \right)
 \right]
 &=&\frac{\omega }{\tau^4} \left(G_{(2)}' +\frac{2\tau}{3 \tau_0^2}\right),
\eea
and for all $n\ge4$:
\bea
\partial_\tau \left[
 \frac{\omega}{\tau^4} \left(\tau_0  \,G_{(n)}' - G_{(n-1)}\right)
 \right]
 &=&
 \frac{\omega\,G_{(n-1)}' }{\tau^4}.
\eea
We follow
 the protocol  of \cite{Tasinato:2020vdk} to
 solve this system in the interval $ \tau_1<\tau<\tau_2$. (We refer
the reader to \cite{Tasinato:2020vdk} for additional technical details.) 
The solution nearby $\tau_1$ is 
\bea
\tau_0^2\,G_{(2)}(\tau)&=&\frac{\alpha}{3} \frac{(\tau-\tau_1)^2}{2},
\\
\tau_0^n\,G_{(n)}(\tau)&=&\frac{\alpha \,\tau_1}{3} \,2^{n-3}\,
\frac{(\tau-\tau_1)^{n-1}}{(n-1)!},
\label{eq_solAN}
\eea
with a constant  parameter $\alpha$ defined as
\be\label{def_of_alp}
\alpha\,=\,\left(\frac{d \ln \omega}{d \ln \tau}\right)_{\big| {\tau=\tau_1}}.
\ee
These solutions correctly describe the system
in the time short interval of interest, $\tau_1\le \tau\le \tau_2$.
The dimensionless quantity $\alpha$ in eq \eqref{def_of_alp} is a key parameter
for us. If  large,  it can considerably amplify the magnetic spectrum from
large towards small scales.

\smallskip

In fact,
plugging the configurations \eqref{eq_solAN} into equation \eqref{eq_ans2} yields a series of the form
\be
{\cal A}_k(\tau)=\frac{3\,a^2\,\sqrt{\omega}
\,H_I^2\,e^{-i k \tau}}{\sqrt{2\,k^5}}\,
\Big[1+ i k \tau - \frac{k^2 \tau^2}{3} - \frac{\alpha}{6}k^2(\tau - \tau_1)^2 - \frac{i \alpha k \tau_1}{12}\Big(1 + 2ik(\tau-\tau_1) - \sum_{n = 0}^\infty \frac{(2ik)^n}{n!}(\tau-\tau_1)^n \Big)\Bigg].
\ee
The exponential series can be resummed, yielding an exact solution
\bea
{\cal A}_k&=& \frac{3\,a^2\,\sqrt{\omega}
\,H_I^2\,e^{-i k \tau}}{\sqrt{2\,k^5}} \left[1+ i k \tau-\frac{k^2 \tau^2}{3} 
-\frac{\alpha}{6} k^2 (\tau-\tau_1)^2-
\frac{i \,\alpha\, k \,\tau_1 }{12}
\left( 1+2 i k (\tau-\tau_1)-
e^{2 i  k (\tau-\tau_1)}
\right)
\right].
\nonumber\\
\label{eq_solt2}
\eea
This is the analytic solution of the mode function in the interval  $\tau_1<\tau<\tau_2$. 
Implementing Israel
junction conditions, the 
  expression \eqref{eq_solt2}
is then  joined to a second
period of standard power-law evolution
for the conformal function $I(\tau)$  for $\tau_2\le \tau \le \tau_R$, with $\tau_R$
being the conformal time when inflation ends. According to eq \eqref{eq_anin}, in this 
interval we recover   $I(\tau)= a^2(\tau)$ --  the profile associated
with a scale invariant magnetic spectrum. Recalling the definition 
\eqref{eq_charsc} of $k_\star$,
 we introduce the convenient dimensionless variables  $\kappa$ and $\Delta \tau$ 
 as 
\bea
\label{eq_dok}
\kappa &=&k/k_\star\,=\, - \tau_1 k, \\
\tau_2-\tau_1 &=& -\tau_1\,\Delta \tau.
\label{eq_dedt}
\eea
The solution for the mode function in the interval
 $\tau_2<\tau<\tau_R$ results
\bea
\label{eq_sollp}
{\cal A}_k&=& \frac{3 \, {a}^2(\tau) \, H_I^2}{2 \, 
k^{5/2}
}\,
\left[
C_1 e^{-i k \tau}
\left(1+i k \tau-\frac{k^2 \tau^2}{3}\right) 
+
C_2 e^{i k \tau}
\left(1-i k \tau-\frac{k^2 \tau^2}{3}\right) 
\right],
\eea
with $C_{1,2}$ fixed by Israel conditions
\bea
\label{eq_sollpA}
C_1&=&
1+2 \Delta \tau \,\alpha\, \frac{    \kappa  \left(2 (\Delta \tau -1)^2 (\Delta \tau +1) \kappa ^2+2 i (\Delta \tau -1) (3 \Delta \tau +1) \kappa -9 \Delta \tau +2\right)-6 i}{8 (\Delta \tau -1)^4 \kappa ^3}
\nonumber
\\
&&+\alpha \frac{e^{2 i \Delta \tau  \kappa } (2 (\Delta \tau -1) \kappa +3 i)+2 \kappa -3 i}{8 (\Delta \tau -1)^4 \kappa ^3},
\\
C_2&=&\frac{\alpha  e^{-i (\Delta \tau -2) \kappa } \left(2 \left(\Delta \tau ^2-1\right) \kappa ^2+i (\Delta \tau  (3 \Delta \tau -5)-4) \kappa +6 \Delta \tau +3\right) \sin (\Delta \tau  \kappa )}{4 (\Delta \tau -1)^4 \kappa ^3}
\nonumber
\\
&&+ \frac{\alpha  e^{-i (\Delta \tau -2) \kappa  }\,
\Delta \tau  (\kappa  (4 i (\Delta \tau -1) \kappa -3 \Delta \tau +9)+6 i) \cos (\Delta \tau  \kappa )
}{4 (\Delta \tau -1)^4 \kappa ^3}.
\label{eq_sollpB}
\eea

The  magnetic field spectrum
is then evaluated at the end of inflation,  $\tau\to\tau_R$, using the definition in eq \eqref{eq_gmf}:
\be
\label{eq_summf}
{\cal P}_B(k)\,=\,\frac{9 H_I^4}{4 \pi^2}
|{ C}_1(k)+{ C}_2(k)|^2,
\ee
neglecting  terms depending on $k \tau_R$, and
assuming that this quantity is small. Eqs \eqref{eq_sollp}-\eqref{eq_summf} 
are
what
 we need for the considerations we develop next.
 At  large scales, $\kappa\to0$, the magnetic spectrum approaches a constant, with an  amplitude corresponding 
to the scale invariant case of eq \eqref{eq_pbst}:
\be
\label{eq_respsl}
{\cal P}_B(\kappa\ll1)\,=\,\frac{9 H_I^4}{4 \pi^2\,
}.
\ee
Given the smallness of $H_I$ in Planck units, this value
easily satisfies CMB bounds. 
It is then  straightforward to compute the amplitude of the spectrum at very small scales, finding
\be
\label{eq_pbsm}
{\cal P}_B(\kappa\gg1)\,=\,\frac{9 H_I^4}{4 \pi^2}
\left[1+ \frac{\alpha\,
\Delta \tau \,(1+\Delta \tau)
\,
\left\{4+\Delta \tau \,\left[\alpha-8+
\Delta \tau (\alpha+4)
\right] \right\}}{4 (1-\Delta \tau)^4} \right].
\ee
So, comparing with eq \eqref{eq_respsl},  we
can obtain  an enhancement in ${\cal P}_B$ controlled by the parameter $\alpha$. Such mechanism might then be able to produce
small-scale magnetic fields compatible with cosmological observations,
while satisfying stringent constraints
at large CMB scales.

The natural question we address next is {\it how fast} can the magnetic spectrum increase
from large (eq \eqref{eq_respsl})
towards small  (eq \eqref{eq_pbsm}) scales in this context? In what comes next, we show that there is a limitation on its growth rate.

\subsection{The maximal growth of the spectrum: analytical results}
\label{sec_anmx}

\begin{figure}[t!]
    \centering
    \includegraphics[width=0.47\linewidth]{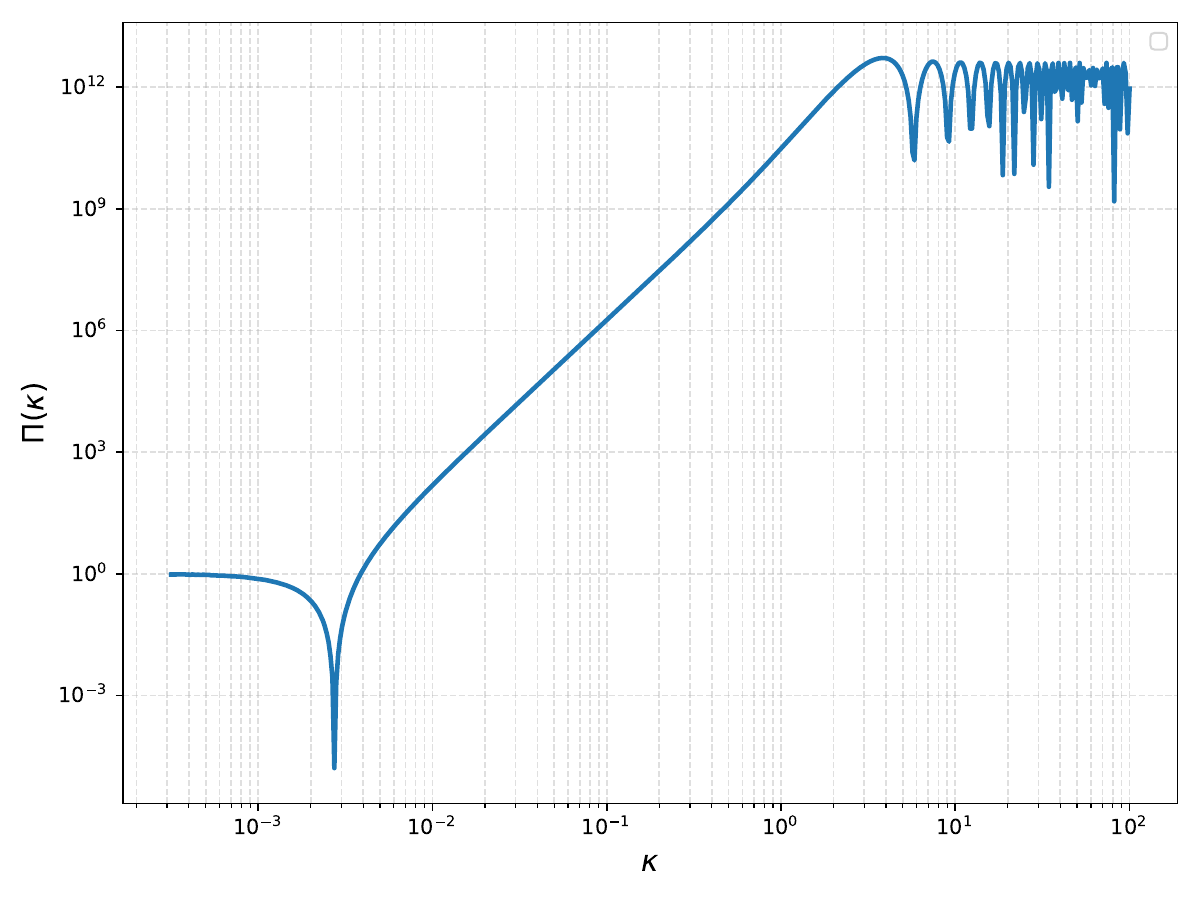}
        \includegraphics[width=0.47\linewidth]{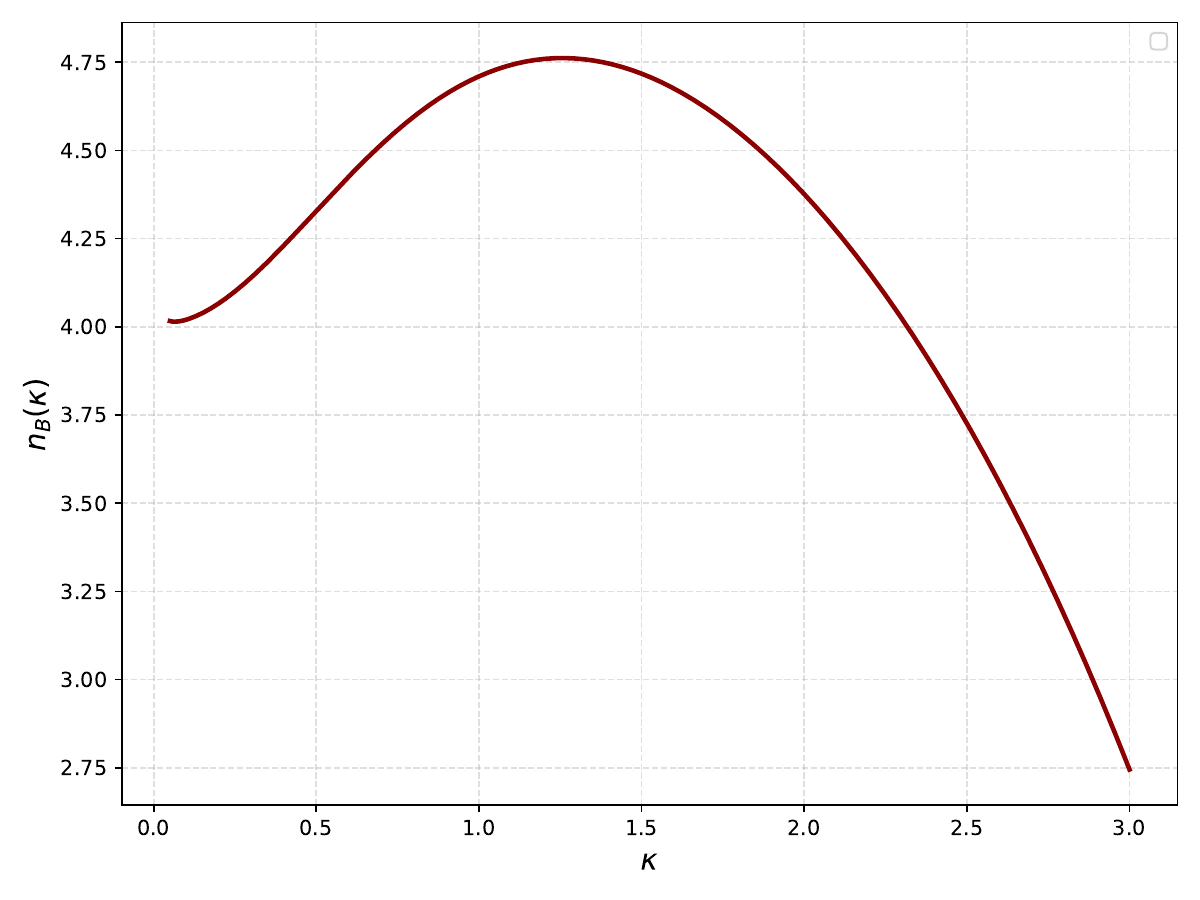}
    \caption{\small {\bf Left panel:}
    Plot of the scale profile of the ratio \eqref{eq_defrat}
    between the magnetic field spectrum 
    against its value at large scales.
    We use the dimensionless variable 
$\kappa$, defined in eq \eqref{eq_dok},
    and  choose $\Pi_0=10^6$. 
    {\bf Right panel:} Plot of the magnetic 
    spectral index $n_B$ as function of the scale, eq \eqref{eq_spectind}, focusing
    on the region of growing magnetic spectrum. The spectral index has a maximum 
    at position $\kappa_{\rm max}\simeq 1.256$ and value
    $n_B(\kappa_{\rm max}) \simeq 4.75$ , after which it decreases to join the flat plateu (on average) towards small scales. 
    }
    \label{fig_growthMF2}
\end{figure}

We  proceed with the characterisation of the  shape  of the magnetic field spectrum in our framework
which includes a brief violation of slow-roll conditions\footnote{Recall the discussion
before eq \eqref{eq_anin}: we denote with `violation of slow-roll' a phase
during which the conformal function $I(\tau)$ is not a simple power law, and obeys
an Ansatz as \eqref{eq_anin}.}. We notice that
considerable simplifications occur in the limit of infinitesimal
$\Delta \tau$ (see the definition \eqref{eq_dedt}), and very large $\alpha$ (see  \eqref{def_of_alp}) --  
yet ensuring 
that their product remains finite
\begin{eqnarray}
&&
\Delta \tau\to0\hskip0.5cm,\hskip0.5cm
\alpha\to\infty
\hskip0.5cm{\rm but}\hskip0.5cm
\alpha \Delta \tau\,=\,2\,\Pi_0 \hskip0.5cm {\text{with\,$\Pi_0$\,\,finite
}}.
\label{eq_laral}
\end{eqnarray}
This limit resembles the `t Hooft limit of large $N$ gauge theories \cite{tHooft:1973alw} (see \cite{Tasinato:2023ukp,Tasinato:2023ioq} for further discussions in a related inflationary  context). 
Then the quantities $C_{1,2}$  \eqref{eq_sollpA}, \eqref{eq_sollpB}  reduce to  simple expressions
\bea
C_1&=&1+ \left(1 - \frac{3 i}{\kappa^3}   - \frac{2 i}{\kappa}  \right) \Pi_0,
\\
C_2&=&-\frac{e^{2 i \kappa } \left(\kappa ^3+4 i \kappa ^2-6 \kappa -3 i\right) \Pi_0}{\kappa ^3}.
\eea
In order to
 study the slope of ${\cal P}_B(\kappa)$  in the general case,
 it is useful to introduce the ratio
\be
\label{eq_defrat}
\Pi(\kappa)\,=\,\frac{{\cal P}_B(\kappa)}{{\cal P}_B(\kappa\ll1)}
\ee 
of the spectrum evaluated   
at  scale 
$\kappa =k/k_\star$ (with $k_\star=-\tau_1$), against its
constant value 
at large scales $\kappa\ll1$. The function $\Pi(\kappa)$ in eq \eqref{eq_defrat} 
singles out the scale dependence of the spectrum. Using formula \eqref{eq_pbsm},
 in the limit of eq \eqref{eq_laral}, this quantity results
\bea
\Pi(\kappa)&=&1+
\frac{2 {\Pi_0} \left(\kappa ^3+\left[4 \kappa ^2-3\right) \sin (2 \kappa )-\left(\kappa ^2-6\right) \kappa  \cos (2 \kappa )\right]}{\kappa ^3}
\nonumber
\\
&&+
\frac{4 \left(\kappa ^2+1\right) {\Pi_0}^2 \left[\left(\kappa ^2-3\right) \sin (\kappa )+3 \kappa  \cos (\kappa )\right]^2}{\kappa ^6}.
\label{eq_sips}
\eea
It  depends on the single parameter $\Pi_0$ as defined in eq \eqref{eq_laral}. 
In Fig \ref{fig_growthMF2}, left panel, 
we represent $\Pi(\kappa)$
 as function
of  the dimensionless quantity $\kappa$.
When
expressed in terms of $\Pi_0$,  the enhancement from
large towards small scales results
\be
\label{eq_finen}
\Pi(\kappa\gg1)\,=\,(1+\Pi_0)^2.
\ee
Since magnetogenesis aims to achieve an enhancement of
the spectrum towards small scales, we select a large value for 
the dimensionless constant $\Pi_0$   (more on this
in section \ref{sec_strcou}, where we show that $\Pi_0$
should be at least of order $10^6$). At very small scales, for $\kappa\gg1$, though,
we expect that the magnetic field amplitude gets damped by dissipation or effects
associated with 
magnetohydrodynamic (MHD) turbulence of the astrophysical plasma: see e.g. \cite{Jedamzik:1996wp}. 
 This suppression will play an important role in the analysis 
 of induced gravitational waves in the next section.

\medskip

The magnetic spectrum profile in the left panel of  Fig \ref{fig_growthMF2}
has a dip at intermediate scales.
Using the same
arguments of \cite{Tasinato:2020vdk,Tasinato:2023ukp},
it is  straightforward to determine
its position  as a function of $\Pi_0$, in the limit of large $\Pi_0$
we are interested in. We find
\be
\kappa_{\rm dip} = \sqrt{\frac{15}{2}}\,\frac{1}{\Pi_0^{1/2}},
\ee
up to corrections that are suppressed by higher powers of $1/\Pi_0$. 
In proceeding from small towards large $\kappa$,
after
the dip the magnetic field spectrum profile $\Pi(\kappa)$  grows steadily, to then 
join  an oscillatory region with approximately constant amplitude  at small scales ($\kappa \simeq 5$).

  Working in the
 limit of large $\Pi_0$, we can determine the spectral index
 associated with the magnetic field spectrum,
 finding
 \bea
n_B(\kappa)&=&\frac{d \ln \Pi(\kappa)}{d \ln k}
\nonumber
\\
&=&\frac{2 \left(-3 \kappa ^4+2 \kappa ^2+9\right) \sin (\kappa )+2 \kappa  \left(\kappa ^4-5 \kappa ^2-9\right) \cos (\kappa )}{\left(\kappa ^2+1\right) \left[\left(\kappa ^2-3\right) \sin (\kappa )+3 \kappa  \cos (\kappa )\right]}
+{\cal O}\left(\frac{1}{\Pi_0}\right).
\label{eq_spectind}
\eea
Relation \eqref{eq_spectind} is what we 
need for our aim to find the maximal slope
of the spectrum. The function
is  easy to handle, and we represent it
in the right panel of Fig \ref{fig_growthMF2}, focusing in the region
of scales where the magnetic  spectrum in the figure left panel  grows. We find
that $n_B(\kappa)$ has a maximum at around 
\be
\label{eq_ktan}
\kappa_{\rm max}\,=\,1.256,
\ee
where 
\be
\label{eq_nmax}
n_{\rm max}^{(B)}=
n_B(\kappa_{\rm max})=4.75.
\ee
Hence, $n_{\rm max}$ is the maximal slope
we can achieve in the specific limit \eqref{eq_laral}
we are considering. 
See also Appendix 
\ref{app_est} for a more systematic analysis away from
the limit \eqref{eq_laral}. 

\smallskip
Eq \eqref{eq_nmax} leads us to conclude that the growth of the magnetic
field spectrum during a phase of non-slow-roll -- starting from a scale-invariant profile at large scales  --
can be steeper with respect to the case of
scalar curvature fluctuations \cite{Byrnes:2018txb}. In the scalar case,  the maximal
slope was found to be $n^{\rm{(scal)}}_{\rm max}\,=\,4$, up to logarithmic corrections \cite{Carrilho:2019oqg,Ozsoy:2019lyy}. The difference among the results is due to the 
structure of  equation \eqref{eq_eom}
with $I(\tau)$  given in eq \eqref{eq_anin},
which is different
with respect 
to the scalar case\footnote{Preliminary investigations of the analogous questions in the spin-2 (tensor) case have been presented in \cite{Mylova:2018yap,Ozsoy:2019slf}.}.
  We now continue with
  a discussion of backreaction and strong coupling issues,
  to then discuss 
 phenomenological consequences of these findings.

\subsection{Some considerations on backreaction and strong coupling issues}
\label{sec_strcou}

In this section we estimate some features of the magnetic field
spectrum produced by the non-slow-roll evolution discussed above, and we address
potential  issues regarding theoretical aspects of our setup.
The amount of growth
in the magnetic field spectrum is governed by the quantity $\Pi_0$, depending
on the slope of the conformal function $I(\tau)$
during the non-slow-roll evolution. 
We can estimate interesting
values of $\Pi_0$ as follows. Astrophysical 
and cosmological observations seem to suggest that magnetic
fields can be found at different
astrophysical scales, with values of the order of few to tens of  micro-Gauss.  In this section we assume that  fields of such size are produced by the primordial process we are considering, and then evolved or maintained by dynamo effects (see the review 
\cite{Widrow:2002ud}).

Expressing
as $\delta_B \simeq {\cal P}_B^{1/2}$ the size of magnetic field at the end of inflation, 
we have  (see eq \eqref{eq_finen})
\be
  \delta_B\,=\,\frac{3H_I^2}{2\pi}(1+\Pi_0),
\ee
once this quantity is evaluated at small  scales $k\ge k_\star$. After inflation ends, this value
is rapidly diluted by cosmological expansion, the suppression factor scaling
as $(a_R/a_0)^{2}\,=\,10^{-29}$ assuming an instantaneous
reheating process \cite{Demozzi:2009fu}. (Here $a_R$ is the scale factor at the end of inflation, or beginning of radiation domination era, while $a_0$ the scale factor today.) Hence, assuming $H_I = 10^{-6}$
in Planck units, and converting the results to Gauss ($1$ Gauss $=1.95\times 10^{-2}$ eV$^2$), we  require
that  the amplitude of the spectrum $\delta^{(0)}_B$  today is
\begin{align}
    \delta^{(0)}_B&=\frac{3H_I^2}{2\pi}(1+\Pi_0)\left(\frac{a_R}{a_0} \right)^2\,=\,\left(\frac{\delta^{\rm obs}_B}{10^{-5}\,\,{\rm G}}\right) 10^{-5}\,\,{\rm G}\\
    &\simeq 10^{46}(1+\Pi_0)(10^{-29})^2\,\,{\rm G}\,=\,\left(\frac{\delta^{\rm obs}_B}{10^{-5}\,\,{\rm G}}\right) 10^{-5}\,\,{\rm G},
\end{align}
where we select a pivot value of $10$ micro-Gauss for selecting a  value of $\delta^{\rm obs}_B$
compatible with astrophysical  observations. 
 {Although this amplitude is larger than the required value of the seed magnetic field -- typically of the order of nano-Gauss -- there are various damping and diffusion mechanisms occurring in the post-inflationary universe which can dissipate the magnetic field energy over the scales of our interest (in this context, see~\cite{Jedamzik:1996wp, Cruz:2023rmo, Paoletti:2022gsn, Chluba:2015lpa}).}
Solving for $\Pi_0$,
we find
\begin{equation}
\label{eq_despi}
    \Pi_0\sim 10^7\,\left(\frac{\delta^{\rm obs}_B}{10^{-5}\,\,{\rm G}}\right).
\end{equation}
Consequently,  typical values for $\Pi_0$ to obtain cosmologically
significant magnetic fields at small scales
span from $10^6$ to few times $10^7$.  Importantly, although these are
the orders of magnitude of $\Pi_0$  preferred by cosmological observations, theoretical
considerations prevent us from choosing $\Pi_0$ much larger than $10^6$ -- $10^7$. In fact, the amplitude of the magnetic field energy at small scales goes as $
\rho_B\sim
{\cal P}_B \,\propto\,H_I^2 \times \left( H_I\,\Pi_0\right)^2
$. To avoid large backreaction on the inflationary dynamics, whose typical
energy scale is of the order $\rho_{\rm inf}\sim H_I^2$, the factor $H_I\,\Pi_0$ can be at most of order one. Hence  choosing $H_I\sim 10^{-6}$ in Planck units, we can not select too large values of $\Pi_0$. 

\medskip

The characteristic comoving momentum \( k_\star = -1/\tau_1 \), around which the rapid growth of the spectrum occurs --- see Fig.~\ref{fig_growthMF2} --- is determined by the  time \( \tau_1 \) at which a brief non-slow-roll phase takes place during inflation. Depending on the location of \( k_\star \), magnetic fields can be amplified on a range of scales, from large   distances corresponding
to galaxy clusters ($10^{24}$ cm) to relatively small stellar ones ($10^{10}$ cm).

\smallskip

Our configuration, in which the gauge kinetic function \( I(\tau) \) undergoes a rapid change during a short interval, resonates with scenarios such as those proposed in~\cite{Ferreira:2013sqa}, where a sawtooth profile for \( I(\tau) \) is shown to mitigate the strong coupling problem in primordial magnetogenesis~\cite{Demozzi:2009fu} (see also~\cite{Tasinato:2014fia} for alternative approaches to this issue).
In essence, the problem is the following: if one chooses a function \( I(\tau) \) that increases during inflation --- as in the scale-invariant case \( I(\tau) = a^2(\tau) \) --- and imposes \( I(\tau_R) = 1 \) at the end of inflation, then inevitably \( I(\tau) \) is extremely small at early times. Since \( I(\tau) \) is inversely proportional to the electromagnetic coupling, this leads to an unphysically large coupling at early times, rendering the electromagnetic theory unreliable.

\smallskip

A proposed resolution~\cite{Ferreira:2013sqa} involves constructing scenarios in which \( I(\tau) \) changes its slope during inflation, keeping its amplitude sufficiently large to avoid strong coupling throughout. In our model, a similar mechanism is potentially operative: by choosing a sufficiently large value of \( \omega_0 \) in eq.~\eqref{ans_ome}, we ensure that \( I(\tau) \) remains large enough at the onset of inflation. Furthermore, if the non-slow-roll phase begins sufficiently late---that is, if \( \tau_1 \) is chosen appropriately---then \( I(\tau) \) remains at safe values thereafter, ensuring theoretical control of the setup\footnote{Alternatively, one might consider multiple short non-slow-roll epochs~\cite{Tasinato:2020vdk}, stitched together by segments in which \( I(\tau) \) follows different power-law behaviors. This would allow one more flexibility in choosing $\tau_1$.}. It would be interesting to further develop these preliminary ideas within a fully developed framework.

\section{Gravitational waves induced by magnetic field amplification}
\label{sec_gwa}

In the previous sections, we learned that a period of non-slow-roll 
evolution can amplify the magnetic field spectrum towards small scales, 
at around a comoving scale $k_\star=-1/\tau_1$  corresponding to the 
brief inflationary epoch during which slow-roll is violated. In this section, we analyse
the properties of Gravitational Waves (GWs) sourced by such an amplified
magnetic field. We show that the frequency dependence of the  induced Stochastic GW Background (SGWB) can constitute a distinct  experimental  smoking gun for this scenario. 
Interestingly, even if the magnetic field is rapidly damped by turbulence or diffusion after its production, the gravitational waves it induces can still serve as evidence of its existence
in the first place. 
Our findings provide specific templates for stochastic gravitational wave background (SGWB) profiles (see, e.g., \cite{Caprini:2024hue,Blanco-Pillado:2024aca,LISACosmologyWorkingGroup:2024hsc}), highlighting inflationary magnetogenesis as a compelling target for gravitational wave searches.

\bigskip

To compute 
the induced GW spectrum,  we follow the methods
developed in 
\cite{Baumann:2007zm,Kohri:2018awv} in
the context of stochastic gravitational
wave backgrounds (SGWB) in scalar-induced  scenarios where
amplified  curvature fluctuations source the GW after inflation ends. This subject has a long
history  -- see e.g. 
\cite{Matarrese:1992rp,Ananda:2006af,Baumann:2007zm,Saito:2009jt,Bugaev:2009zh,Espinosa:2018eve,Kohri:2018awv},  and  \cite{Domenech:2021ztg} for a comprehensive review.
We apply the idea
to the non-adiabatic case of magnetic field sources, a topic 
 studied
in several works \cite{Durrer:1999bk, Caprini:2001nb,Mack:2001gc,  Pogosian:2001np, Caprini:2003vc, Shaw:2009nf, Saga:2018ont,Bhaumik:2025kuj, Maiti:2025cbi},
 although usually  focusing 
on simple power law profiles
for the magnetic field correlator ${\cal P}_B$ in eq \eqref{eq_gmf}. Here we 
consider much richer scale-dependent profiles for ${\cal P}_B$,   as motivated
by the considerations of section \ref{sec_maxs} on non-slow-roll evolution during inflation
-- hence we need to further develop
the corresponding formalism for GW production.

\subsection{The calculation of the gravitational wave spectrum}

After inflation ends, the GW  equation of motion reads
\be\label{eq_strheq}
h_{ij}''(\tau, {\bf x})+2 {\cal H}(\tau)\,h_{ij}'(\tau, {\bf x})-\nabla^2 h_{ij}(\tau, {\bf x})
\,=\,\Pi_{ij}^{(T)}(\tau, {\bf x}),
 \ee
with $\Pi_{ij}^{(T)}$ being the transverse-traceless component 
of the magnetic field stress tensor  sourcing the GW. 
 In order to
express this quantity and proceed with our discussion, it is convenient to work in
Fourier space.
The spin-2 GW fluctuations are decomposed as
 \bea
h_{ij} (\tau, {\bf x})&=&
\sum_{\lambda}
\,\int \frac{d^3 {\bf k}}{(2 \pi)^{3/2}} \,e^{i {\bf k}  \cdot {\bf x} }\,
{\bf e}^{(\lambda)}_{ij}({\bf k})\,h_{\bf k}^{(\lambda)} \,,
\eea
with ${\bf e}^{(\lambda)}_{ij}({\bf k})$
the spin-2 polarisation tensors. The evolution equation for the modes $h_{\bf k}^{(\lambda)}$ results
\be
\label{eq_eveqh}
h_{\bf k}^{(\lambda)''} +2 {\cal H}\,h_{\bf k}^{(\lambda)'}
+k^2 \,h_{\bf k}^{(\lambda)}
\,=\,S^{(\lambda)}(\tau, {\bf k}).
\ee
The source  in the right hand
side of this equation reads
\bea
\label{eq_gwsor}
S^{(\lambda)}(\tau, {\bf k})&=&{\bf e}^{(\lambda)\,ij}(\hat k) \,
\Pi_{ij}^{(T)}(\tau, {\bf k})
\,=\,
\frac{2\,{\bf e}^{(\lambda)\,ij}(\hat k)
\,\Lambda_{ij}^{\,\,\,\,mn}
\tau_{mn}^{(B)}({\bf k})}{a^2(\tau)},
\eea
where the magnetic field stress tensor is (see e.g. 
\cite{Mack:2001gc}) 
\be
\label{eq_emtmf}
\tau_{ij}^{(B)}({\bf k})
\,=\,\frac{1}{4 \pi}\int \frac{d^3 p}{(2\pi)^3}
\left[B_i({\bf p})B_j({\bf k}-{\bf p})-\frac{\delta_{ij}}{2}
B_m({\bf p})B_m({\bf k}-{\bf p})
\right].
\ee
The projection tensor $\Lambda_{ij}^{\,\,\,\,mn}$ selects its transverse-traceless
part, and is given by

\be
\Lambda_{ij}^{\,\,\,\,\ell m}\,=\,\frac12
\left(\pi_i^\ell\,\pi_j^m+\pi_j^\ell\,\pi_i^m
-\pi_{ij} \pi^{\ell m}\right)\,, \hskip0.7cm {\rm with} \hskip0.7cm
 \pi_{ij}=\delta_{ij}-\hat k_i \hat k_j.
\ee
Notice that $\Lambda_{ii}^{\,\,\,\,\ell n}=\Lambda_{ij}^{\,\,\,\,\ell \ell}=0$. Hence  we can neglect the contribution proportional to
$\delta_{ij}$  in \eqref{eq_emtmf}. The quantities entering eq \eqref{eq_emtmf} are evaluated
at the end of inflation. Their value then redshifts with the universe 
expansion after inflation ends, as indicated by the scale factor dependence
of the source \eqref{eq_gwsor}.

 \smallskip
 Equation \eqref{eq_eveqh} can be formally solved as
\be
\label{eq_frh}
h^{(\lambda)}_{\bf k}(\tau)\,=\,\frac{1}{a(\tau)}\,\int d \tau' g_{k}(\tau, \tau') \left[ a(\tau')\,S^{(\lambda)}_{\bf k}(\tau')
\right]\,,
\ee
where the $g_k$ is the Green function evaluated
at the epoch of interest. During radiation domination -- the era on which we focus
our attention from now on -- it reads
\be
g_{k}(\tau, \tau')\,=\,\frac{1}{k} \left[ \sin{(k \tau)} \cos{(k \tau')}-\sin{(k \tau')}
\cos{(k \tau)}
\right]\,.
\ee
The result \eqref{eq_frh} allows us
to formally express the tensor power spectrum
as
\bea
{\cal P}_h(\tau, k)
&\equiv&\frac12\,\frac{k^3}{2 \pi^2}
\,\sum_\lambda
\langle
h^{(\lambda)}_{\bf k}(\tau) 
h^{(\lambda)\,\,*}_{\bf q}(\tau) 
\rangle'_{\bf k=\bf q}
\\
&=&\frac{ k^3}{4 \pi^2\,a^2(\tau)} 
\int d \tau_1 d \tau_2\,g_{ k}(\tau, \tau_1)\,g_{ q}(\tau, \tau_2)\,a(\tau_1) a(\tau_2) \left(\sum_\lambda
\langle
S^{(\lambda)}_{\bf k}(\tau_1)
S^{(\lambda)\,\,*}_{\bf q}(\tau_2)
\rangle'_{\bf k=\bf q} \right)\,,  
\nonumber\\
\label{eq_ts2s}
\eea
where as before the symbol $\langle\dots \rangle'_{\bf k=\bf q}$ indicates two-point correlators
omitting the momentum-conserving $\delta-$function. 
The GW spectrum above is
 the necessary ingredient to compute the GW 
 density parameter, $\Omega_{\rm GW}$, which is the basic quantity 
 to be compared with experiments. Following the notation 
 of \cite{Kohri:2018awv}, we have
\bea
\label{def_ogw}
\Omega_{\rm GW}&\equiv&
 \frac{k^2}{12\,a^2 H^2}
\bar{{\cal P}}_h,
\eea
where a bar indicates average over rapid oscillations.
To proceed, we need to estimate $\bar{{\cal P}}_h$. We  compute
the quantity within parenthesis in eq
\eqref{eq_ts2s}:
\bea
\sum_\lambda
\langle
S^{(\lambda)}_{\bf k}(\tau_1)
S^{(\lambda)\,*}_{\bf q}(\tau_2)
\rangle'_{\bf q=\bf k}
&=&\frac{4\,\Lambda_{ij}^{\,\,\,\,\ell n}}{a^2(\tau_1) \,a^2(\tau_2)}
\,\langle \hat\tau^{(B)\,\,ij}({\bf k}) \hat \tau_{\ell n}^{(B)\,*}({\bf p})  \rangle'_{{\bf p}={\bf k}} 
\\
&=&\frac{4}{(4 \pi)^2\,a^2(\tau_1) \,a^2(\tau_2)}
\nonumber
\\
&\times&
\int \frac{d^3 p_1}{(2\pi)^3}
\,\Lambda_{ij}^{\,\,\,\,\ell n}(\hat k)\,\left( 
\pi^i_\ell(\hat p_1) \pi^j_n(\hat n) +
\pi^i_n(\hat p_1) \pi^j_\ell(\hat n) 
\right)\,P_B(p_1) P_B(|{\bf k}-{\bf p}_1|),
\nonumber\\
\label{eq_sumS}
\eea
where we introduce the unit tensor $\hat n= ({\bf k}-{\bf p}_1)/|{\bf k}-{\bf p}_1|$, and 
the amplitude of the magnetic field 2-point correlators is introduced
in eq \eqref{eq_mfa}.   
We do not consider non-Gaussian contributions, such as those arising from connected four-point functions of the magnetic field, since the underlying Maxwell action is quadratic in the vector fields and therefore does not generate intrinsic non-Gaussianity. 
Following
 \cite{Baumann:2007zm,Kohri:2018awv},
 we introduce
 convenient variables
\be
 u\equiv \frac{
|{\bf k}-{\bf p}_1|}{k}\hskip0.5cm,\hskip0.5cm v\equiv \frac{p_1}{k}
\hskip0.5cm,\hskip0.5cm
\mu \equiv \frac{{\bf k} \cdot {\bf p}_1}{k \,p_1}
\,=\, \frac{1-u^2-v^2}{2 v}.
\ee
 The tensors within the integral of eq \eqref{eq_sumS} can be contracted
 straightforwardly \cite{Caprini:2003vc}, leading to a function of $(u, v)$ which we call $C_0$
 \bea
{\cal C}_0(u,v)&=& \Lambda_{ij}^{\,\,\,\,\ell n}(\hat k)\left( 
\pi^i_\ell(\hat p_1) \pi^j_n(\hat n) +
\pi^i_n(\hat p_1) \pi^j_\ell(\hat n) 
\right)
\\
&=&(1+ \mu^2)\left(1+\frac{ (1-\mu v)^2}{u^2}\right).
\eea
By expressing $d^3 p_1 = 2 \pi p_1^2 d p_1 d \mu$, we can re-write \eqref{eq_sumS} as
\bea
\sum_\lambda
\langle
S^{(\lambda)}_{\bf k}(\tau_1)
S^{(\lambda)}_{\bf q}(\tau_2)
\rangle'
&=&
\frac{ k^3}{4 \pi^2\,a^2(\tau_1)\,a^2(\tau_2)}
\int_0^{\infty} v^2 {d v} \int_{-1}^1 d \mu
\,{  P}_B(k u)
\,{  P}_B(k v)\,{\cal C}_0(u,v)
\nonumber
\\
&=&\frac{\pi^2\,k^{-3}}{a^2(\tau_1)\,a^2(\tau_2)}
\int_0^{\infty} \frac{d v}{u^3\,v} \int_{-1}^1 d \mu
\,{ \cal P}_B(k u)
\,{\cal  P}_B(k v)\,{\cal C}_0(u,v),
\label{eq_sigV2}
\eea
where we used eq \eqref{def_ampbs} to pass from ${P}_B$ to ${\cal P}_B$ between the
first and the second line. Eq
\eqref{eq_sigV2} can  be plugged into the definition of ${\cal P}_h$ in \eqref{eq_ts2s}. The tensor
spectrum is then nicely factorised into two integral 
contributions. They are
\bea
{\cal P}_h\,=\,\frac{1}{4\,a^2(\tau)}\,{\cal I}_\tau^2\,{\cal I}_{uv},
\eea
with
\bea
{\cal I}^2_\tau&=&
\left(
\int_{\tau_R}^\tau d \tau_1 \,\frac{g_{ k}(\tau, \tau_1)}{a(\tau_1)} 
\right)^2,
\label{def_IT}
\\
{\cal I}_{uv}&=&\int_0^{\infty} \frac{d v}{u^3\,v} \int_{-1}^1 d \mu
\,{ \cal P}_B(k u)
\,{\cal  P}_B(k v)\,{\cal C}_0(u,v).
\label{eq_iuv}
\eea
We start handling the time integral in eq 
\eqref{def_IT}. Working
in radiation domination, we use the identities $a(\tau )/a(\tau_1)=\tau/\tau_1$ and
$a(\tau )H(\tau)=1/\tau$.
 A simple calculation, averaging over rapid oscillations, gives
at large $\tau$ 
\bea
\bar{ \cal I}_{\tau}^2
&=&\frac{1 }{2\,k^2\,a^4 H^2}
 \left[ \text{Ci}(-k {\tau_R})^2+\left( \frac{\pi}{2} -\text{Si}(-k {\tau_R})\right)^2\right],
\eea
where $\text{Ci}(x)$, $\text{Si}(x)$  are
the cosine and sine integral functions, and with the
bar we average over rapid oscillations.
The integral in eq \eqref{eq_iuv} is  conveniently expressed
in terms of variables $t,s$
\bea
u&=&\frac{t+s+1}{2}
\hskip0.5cm,\hskip0.5cm
v\,=\,\frac{t-s+1}{2}.
\eea
Taking into account the corresponding  Jacobian,
we can  write it as 
\bea
{\cal I}_{uv}
&=&
\int_0^{\infty} d t \int_{-1}^1 d s\,(1-s+t)^{-2}\,
(1+s+t)^{-2}\,
\,{ \cal P}_B(k u)
\,{ \cal P}_B(k v)\,{\cal C}_0(t,s),
\eea
with
\bea
{\cal C}_0(t,s)&=&\left(\frac{\left(s^2+t (t+2)-1\right)^2}{4 (-s+t+1)^2}+1\right) \left(\frac{\left(s^2+t (t+2)+3\right)^2}{4 (s+t+1)^2}+1\right).
\eea
We can pass from the tensor
spectrum to the GW density parameter $\Omega_{\rm GW}$, given by
eq \eqref{def_ogw}. Using our formulas we obtain:
\bea
\Omega_{\rm GW}&=&
\frac{k^2}{96\,a^4\,H^2}\,\bar{\cal I}_\tau^2\,{\cal I}_{uv}.
\label{eq_ts2sV2}
\eea
We 
substitute into eq \eqref{eq_ts2sV2}  the  results derived
above. We use the same expression of \cite{Caprini:2001nb}
\be
 a(\tau)\,=\,H_0 \,\sqrt{\Omega_{\rm rd}}\,\tau,
\ee
for the scale factor during radiation domination, with $H_0$
and $\Omega_{\rm rd}$ respectively
the   Hubble parameter and the fraction
of total energy density in radiation.    We denote with $\hat{\Omega}_B= \rho_B/\rho_{\rm cr}={\cal P}^{\rm CMB}_B/(3 H_0^2)$ a quantity parameterising  the fractional
energy density of the magnetic field at very large
CMB scales, and we use the expression \eqref{eq_respsl} for ${\cal P}_B^{\rm CMB}$.

Finally, 
we multiply the  resulting
 $\Omega_{\rm GW}$
by the factor $\Omega_{\rm rd}$, in order 
 to take into account  the redshift  of the GW observable
 from early to late times 
 (see e.g. 
 \cite{Domenech:2021ztg}). 
Expressing  the formulas in terms of the dimensionless $\kappa$ (see eq \eqref{eq_dok}),  we get
\bea
\label{eq_fiogw}
\Omega_{\rm GW}(\kappa) &=&
\left[\frac{3\,\hat{\Omega}_{B}^2}{64\,\Omega_{\rm rd}}
 \right]
\times\left\{
  \left(  \text{Ci}^2(\kappa\, x_\star) +\left( \frac{\pi}{2} -\text{Si}(\kappa\, x_\star)\right)^2\right) \,\tilde{\cal I}_{uv}(\kappa)
 \right\},
 \eea
 with
 \bea
 \label{eq_tiuv}
 \tilde{\cal I}_{uv}(\kappa)
 &=&\int_0^{\infty} d t \int_{-1}^1 d s\,(1-s+t)^{-2}\,
(1+s+t)^{-2}\,
\,{ \Pi}(\kappa u)
\,{ \Pi}(\kappa v)\,{\cal C}_0(t,s).
 \eea
 The function ${ \Pi}(x)$ is  defined
in eq \eqref{eq_sips}, while we introduce the (small)
number
$
 x_\star \,=\,{\tau_R}/{\tau_1}
 $
which controls the ratio between the time when
inflation ends versus the time $|\tau_1|$ when the slow-roll conditions are violated during inflation. This quantity only enters in the arguments of the $\text{Ci}$, $\text{Si}$ functions.  For clarity, we choose $x_\star = 10^{-4}$, but the results are in any case
mildly dependent on this quantity since, when small, 
it enters only logarithmically in eq \eqref{eq_fiogw} (through $\text{Ci}(x)\sim \ln{(x)}$ for very small $x$). 

\medskip

The overall constant quantity $\left[\dots \right]$ within square brackets in
eq \eqref{eq_fiogw}  depends on the scale of inflation, as
well as on post-inflationary evolution.
It is the same overall constant scale found in previous works
(starting with \cite{Caprini:2001nb}) which quantifies
the impact of the amplitude of large scale magnetic fields 
into GW observables. The overall amplitude is not our primary focus here; instead, we are interested in the scale dependence of the gravitational wave background. For definiteness, we fix the overall prefactor in Eq.~\eqref{eq_fiogw} to a small value, \(\left[ \dots \right] \simeq 2 \times 10^{-48}\), to better highlight the subsequent growth of the spectrum toward smaller scales. Such reduced values
can be due by post-inflationary processes related with field evolution and dissipation
in the astrophysical plasma -- a subject that we do not touch here, though. 

The quantity within curly parenthesis $\left\{\dots \right\}$ 
in eq \eqref{eq_fiogw} is dimensionless, and depends 
on the function $\Pi(\kappa)$ which --
as we learned in the previous section -- controls the growth
of the magnetic field spectrum. The combination within $\left\{\dots \right\}$ can be evaluated
numerically: we find that it has a profile with a plateau, and a maximal value scaling with $\Pi_0$ as 
\be
 \label{eq_esmaa}
 \{\dots \}_{\rm max}\,\simeq\,10^5\,\Pi_0^4.
 \ee

\begin{figure}[t]
    \centering
    \includegraphics[width=0.5\linewidth]{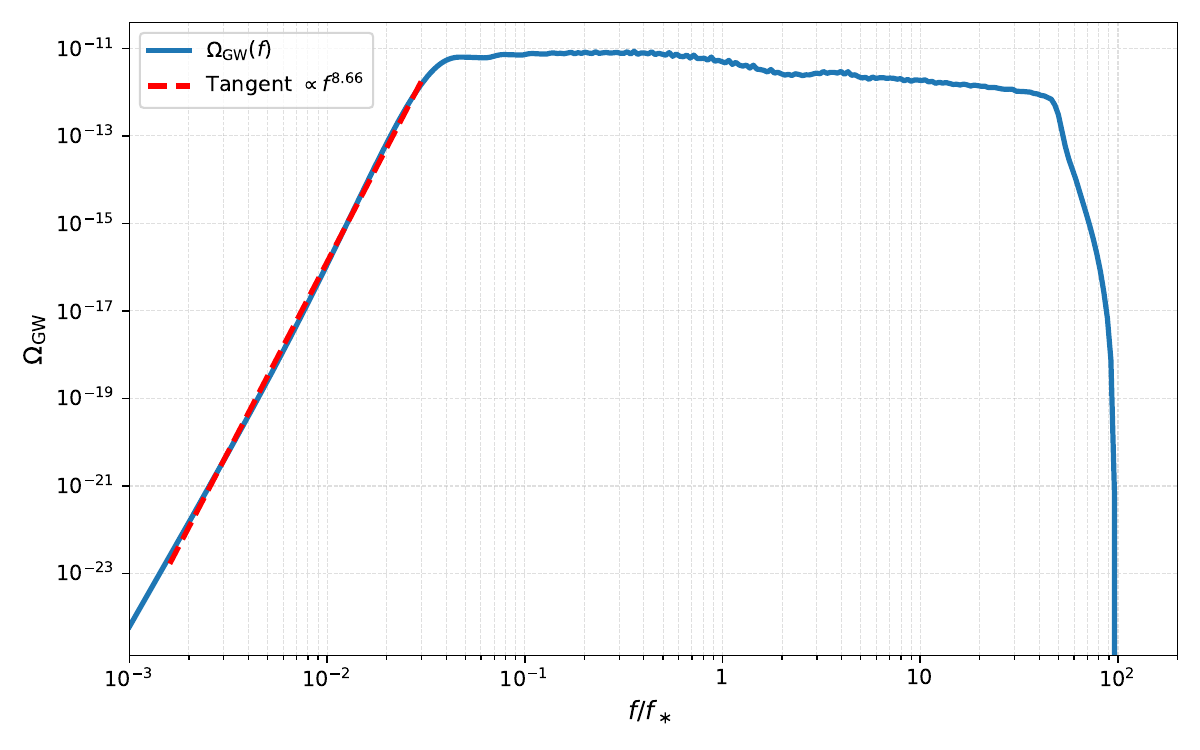}
    \caption{\small 
    Plot of $\Omega_{\rm GW}(f)$ in our
    setup. We follow eq \eqref{eq_fiogw} and choose the parameters as explained in the main text.}
    \label{fig_growthMFv3} 
\end{figure}

Using this information, we plot in Fig.~\ref{fig_growthMFv3} the resulting expression for the fractional energy density \(\Omega_{\rm GW}\) in gravitational waves as a function of frequency, employing the relation \({k}/{\mathrm{Mpc}^{-1}} \simeq 6.5\times 10^{14} \, f/\mathrm{Hz}\). We choose \(\Pi_0 = 7\times 10^7\), a 
large value consistent with our considerations in  section \ref{sec_anmx}.  
To account for the damping of the magnetic field at small scales,
as anticipated in section \ref{sec_maxs}, 
we truncate the magnetic power spectrum at large \(\kappa\), where its amplitude is known to be suppressed by dissipation and turbulent effects (see, e.g., the review~\cite{Durrer:2013pga}).
Specifically, as a concrete example, we use the expression for \(\Pi(\kappa)\) from Eq.~\eqref{eq_sips}, but set \(\Pi(\kappa) = 0\) for \(\kappa > 50\).

\medskip

The spectral shape of the stochastic gravitational wave background (SGWB), \(\Omega_{\rm GW}(f)\), pleasantly  reflects the underlying profile of the magnetic field power spectrum \(\Pi(\kappa)\) that sources it:
\begin{itemize}
\item[-] At relatively low frequencies, \(\Omega_{\rm GW}(f)\) exhibits a steep rise, scaling approximately as \((f / f_\star)^{8.7}\). This behavior arises from the convolution integrals
of eq \eqref{eq_tiuv}, involving the square of the magnetic field spectrum. Since \(\Pi(\kappa)\) during its growth scales with a spectral slope of order 4 (or slightly larger), taking its square within the integrals leads to the aforementioned scaling in the gravitational wave signal.
\item[-]
This rapid growth is abruptly halted at around $f/f_\star\sim 1/50$, giving way to a flat plateau in the SGWB spectrum of Fig \ref{fig_growthMFv3}. The amplitude of this plateau scales as \(\Pi_0^4\).  It 
corresponds to the convolution of the nearly flat region of the magnetic field spectrum at small scales, which follows the initial growth phase -- modulo oscillations in the magnetic field spectrum, which are smoothed out in the convolution integrals.  
\item[-] 
At higher frequencies, specifically for \(f / f_\star \gtrsim 10^2\), the SGWB spectrum undergoes a sharp decline. This reflects the assumed rapid suppression of \(\Pi(\kappa)\) at large \(\kappa \ge 50\), due to dissipation and turbulence effects in the plasma (see discussion following eq \eqref{eq_esmaa}). We emphasise though that this decline in \(\Omega_{\rm GW}(f)\) is  model dependent, since it relies on our assumptions about the suppression 
of the magnetic field spectrum towards small scales.
\end{itemize}

\subsection{Consequences for gravitational wave experiments}

The selected value of the parameter \(\Pi_0 = 7 \times 10^7\)  
 leads to a gravitational wave signal with peak amplitude \(\Omega_{\rm GW} \simeq 10^{-11}\), given our hypothesis
on the overall constant factor in eq \eqref{eq_fiogw}.

\begin{figure}[t]
    \centering
    \includegraphics[width=0.5\linewidth]{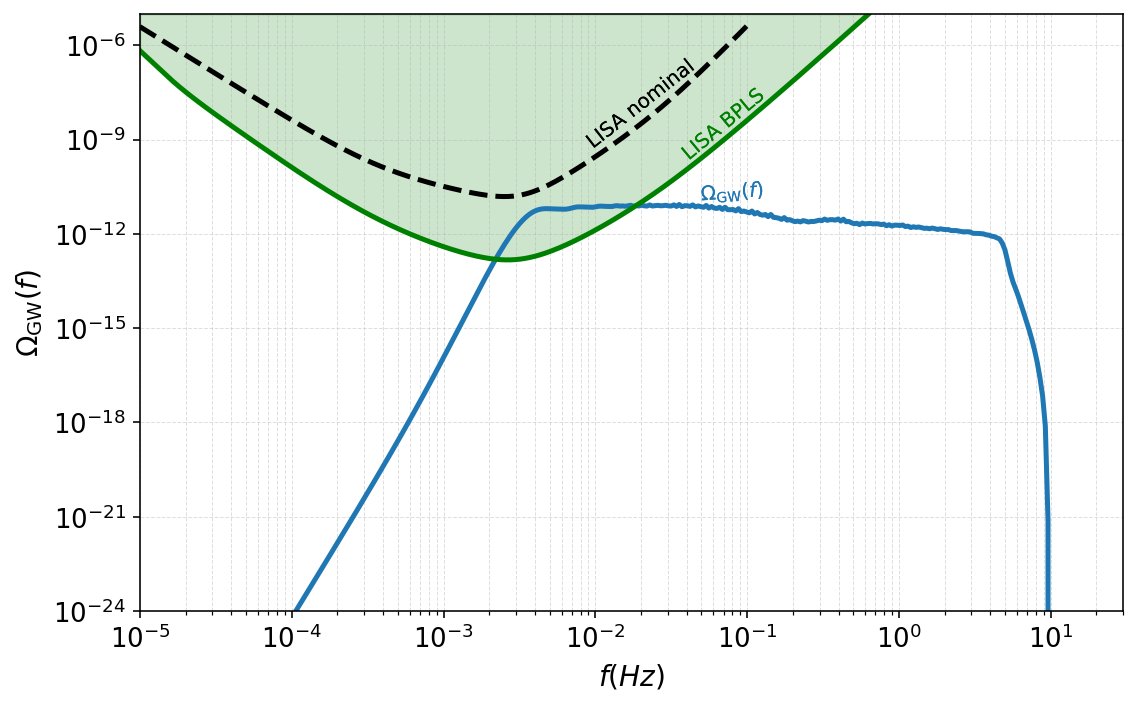}
        \includegraphics[width=0.48\linewidth]{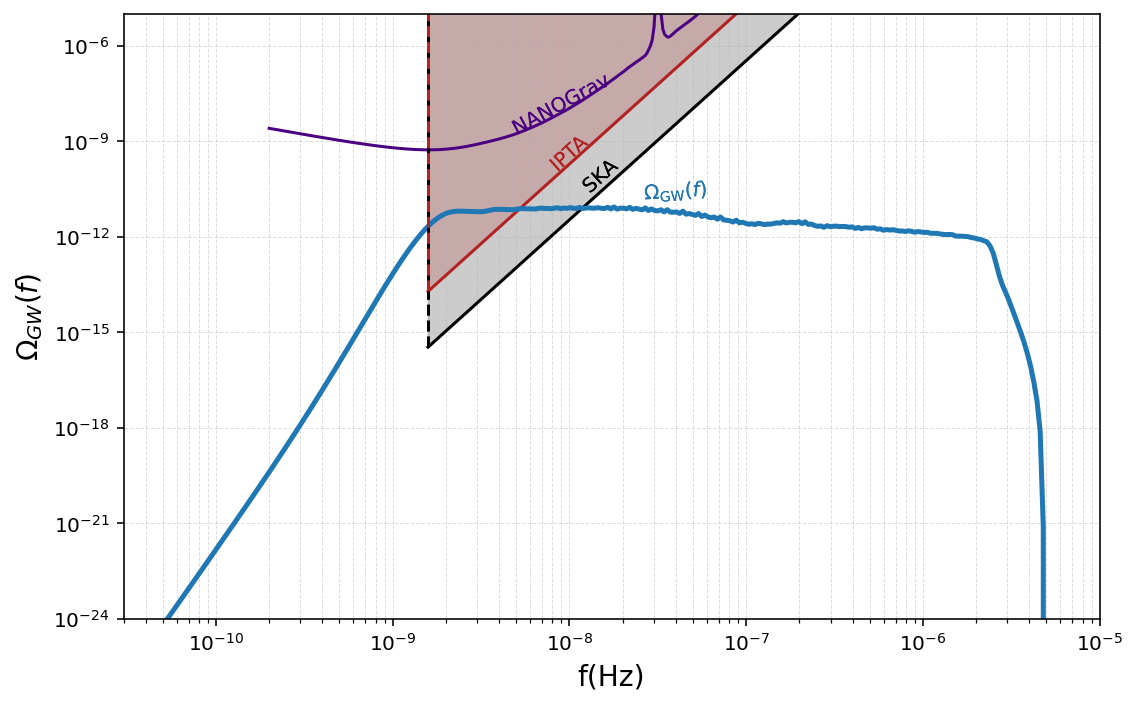}
    \caption{\small {\bf Left:} 
    The same frequency profile of $\Omega_{\rm GW}(f)$ as in Fig.~\ref{fig_growthMFv3} is shown for $f_\star = 0.1$, compared against the nominal LISA sensitivity curve and its broken power-law version \cite{Chowdhury:2022gdc,Marriott-Best:2024anh}.
{\bf Right:}    
    The same frequency profile of $\Omega_{\rm GW}(f)$ as in Fig.~\ref{fig_growthMFv3} is shown for $f_\star = 5\times10^{-8}$, compared against the SKA, IPTA, NANOGrav sensitivity curves. 
    }
    \label{fig_growthMFv7} 
\end{figure}

This GW signal could fall within the sensitivity range of future
GW  experiments provided that the  frequency \(f_\star\)
 determining the position of the plateau in $\Omega_{\rm GW}$ 
lies within their observational bands.
 As specific simple  examples, 
we show in the left panel of Fig \ref{fig_growthMFv7},
that such GW spectrum  can be in principle detected with LISA \cite{LISA:2024hlh}, by choosing  the pivot frequency
$f_\star=0.1$ Hz. Such pivot frequency corresponds
to a magnetic field enhanced at very small scales of $3\times 10^{11}$
cm, i.e. stellar-size scales. In the right panel of Fig \ref{fig_growthMFv7} we instead consider a pivot frequency 
$f_\star=5\times 10^{-8}$ Hz, corresponding to signals at nano-Hertz
scales detectable with pulsar timing array experiments --
we take the corresponding sensitivity curves from \cite{Schmitz:2020syl}. The magnetic
field gets then enhanced a scales of $6\times 10^{17}$
cm, i.e. interstellar size.

The resulting SGWB spectrum has a distinctive shape: 
its rapid growth towards its plateau, and squared shape  is quite atypical compared to  standard templates commonly considered in scalar-induced SGWB scenarios~\cite{LISACosmologyWorkingGroup:2024hsc} based on adiabatic perturbations%
\footnote{
In particular, the logarithmic slope \(\ln(f/f_\star) \sim 8\) characterises the spectral shape in the intermediate regime approaching the plateau, whereas in the deep infrared, the spectrum exhibits a gentler power-law behavior \cite{Cai:2019cdl}.
}. The characteristic knee in the profile of Fig \ref{fig_growthMFv7} can
be detected implementing techniques as \cite{Caprini:2019pxz,LISACosmologyWorkingGroup:2022jok}.
Despite the differences between adiabatic
and non-adiabatic sources, the properties of the magnetic field source can, in principle, be reconstructed using recently developed tools such as those presented in~\cite{LISACosmologyWorkingGroup:2025vdz,Ghaleb:2025xqn}. A more detailed analysis of the detectability and characterisation of the magnetically induced SGWB will be pursued in future work:
nevertheless our investigation already shows  how the rich scale dependence
of the magnetic field spectrum  in our setup leads to a distinctive GW signal.

\section{Conclusions}
\label{sec_out}

We analytically investigated inflationary magnetogenesis in the Ratra model, focusing on scenarios where a brief violation of slow-roll conditions enhances the coupling between the inflaton and gauge fields. This mechanism allows for a rapid growth of the magnetic field spectrum, with an analytically derived maximal slope of \( d \ln {\cal P}_B / d \ln k = 4.75 \), sufficient to bridge the gap between CMB-constrained large-scale amplitudes and observed astrophysical field strengths. We also commented
on strong coupling and backreaction 
problems in this set-up. 

We then studied the stochastic gravitational wave background induced by the amplified magnetic fields, extending standard formalisms to account for the nontrivial spectral features of our scenario. Under suitable conditions, the resulting gravitational wave signal exhibits a characteristic frequency profile and potentially detectable amplitude, providing a unique observational handle on inflationary magnetogenesis with transient non-slow-roll dynamics.

At the technical level, the main highlights of our results are:
\begin{itemize}
\item We introduced a systematic method for analytically studying the spectral profile of perturbations for fields with spin greater than zero, and applied it to the vector case. This method yields compact analytical expressions and allows us to extract key features of the spectrum, such as the location of the dip and the rate of its growth.
\item We derived general expressions for the gravitational wave spectrum induced at second order by magnetic fields, applicable to scenarios with s magnetic field profiles like ours. Remarkably, the resulting formulas exhibit a simpler, factorisable structure than their scalar, adiabatic counterparts—a feature that may prove useful for future analyses of related scenarios.
\end{itemize}

Our findings suggest a  possible connection between primordial magnetogenesis, primordial black hole phenomenology, and gravitational wave physics, motivating further exploration of inflationary scenarios beyond slow-roll and their couplings
with vector fields.

\subsection*{Acknowledgments}
It is a pleasure to thank Pritha Bari,  Anish Ghoshal, Lorenzo Giombi,    Germano Nardini, Marco Peloso and H.V. Ragavendra
 for useful discussions.
We are partially funded by the STFC grants ST/T000813/1 and ST/X000648/1.
For the purpose of open access, the authors have applied a Creative Commons Attribution licence to any Author Accepted Manuscript version arising. Research Data Access Statement: No new data were generated for this manuscript.

\begin{appendix}

\section{ Finite duration of the non-slow-roll epoch}
\label{app_est}

\begin{figure}[h!]
    \centering    \includegraphics[width=0.5\linewidth]{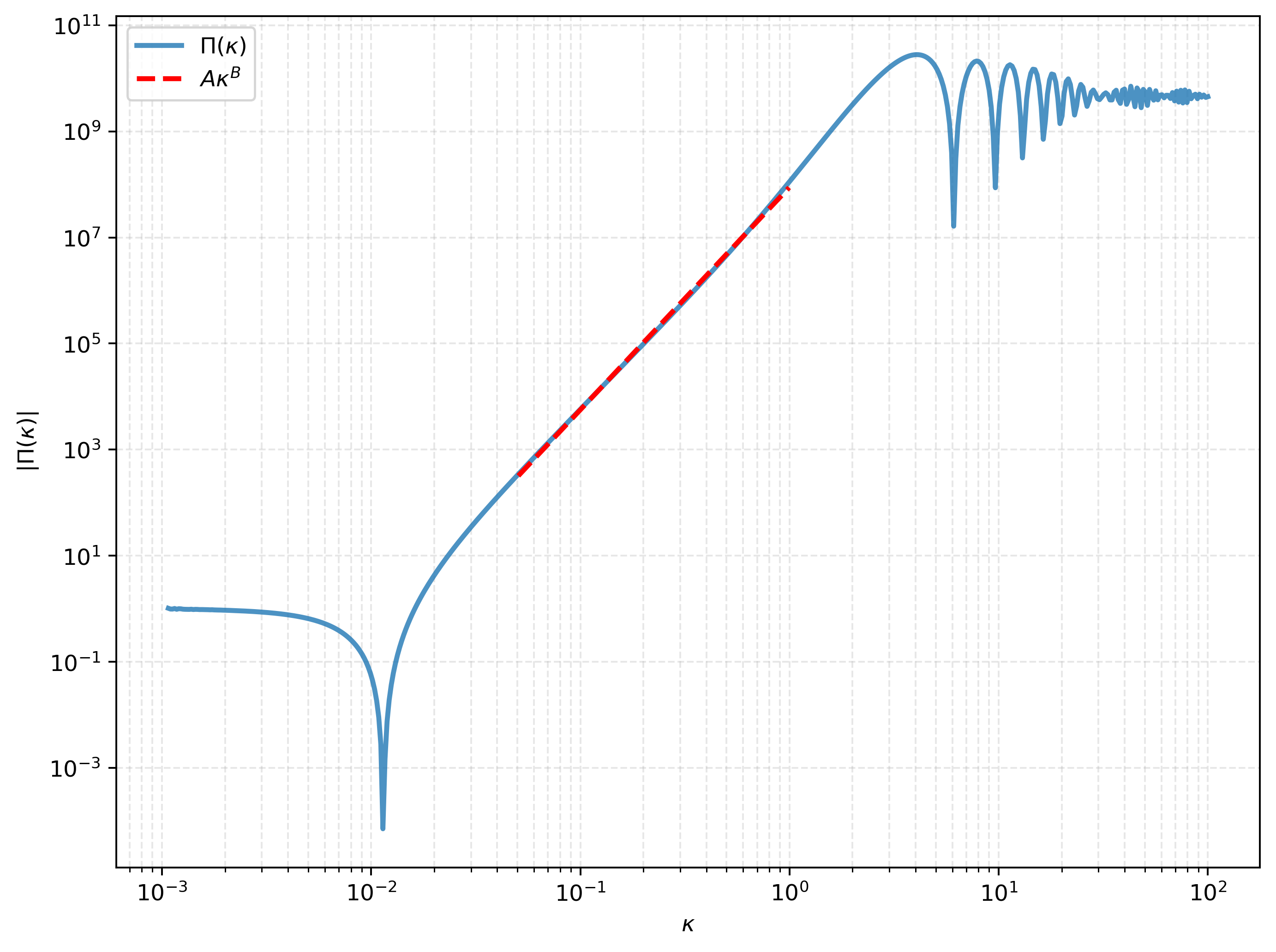}
    \caption{\small An example of the fitted spectrum, $\Pi(\kappa)$, taking $\alpha = 10^7$, $\Delta \tau = 0.1$. The power-law fit finds $A = 8.9 \times 10^7$ $B = 4.197$.}
    \label{powerlaw}
\end{figure}

In this  appendix we reconsider
the problem of determining the maximal
slope of the magnetic field spectrum
in its growing region, without taking
the limit of  eq \eqref{eq_laral}. In other
words, we vary both parameters $\alpha$
and $\Delta \tau$,
  performing a grid-search, computing the maximal slope of $\Pi(\kappa)$ for choices of $\alpha$ and $\Delta \tau$ restricting such that $10^4 \leq \alpha \Delta \tau \leq 10^8$.
  In the growing region we assume the spectrum to follow a power-law profile $A\kappa^B$. Figure \ref{powerlaw} explicitly shows the region in which the power law is fitted, away from deviations from linearity towards the extreme ends of the transitionary phase.

\begin{figure}[h!]
    \centering    \includegraphics[width=0.5\linewidth]{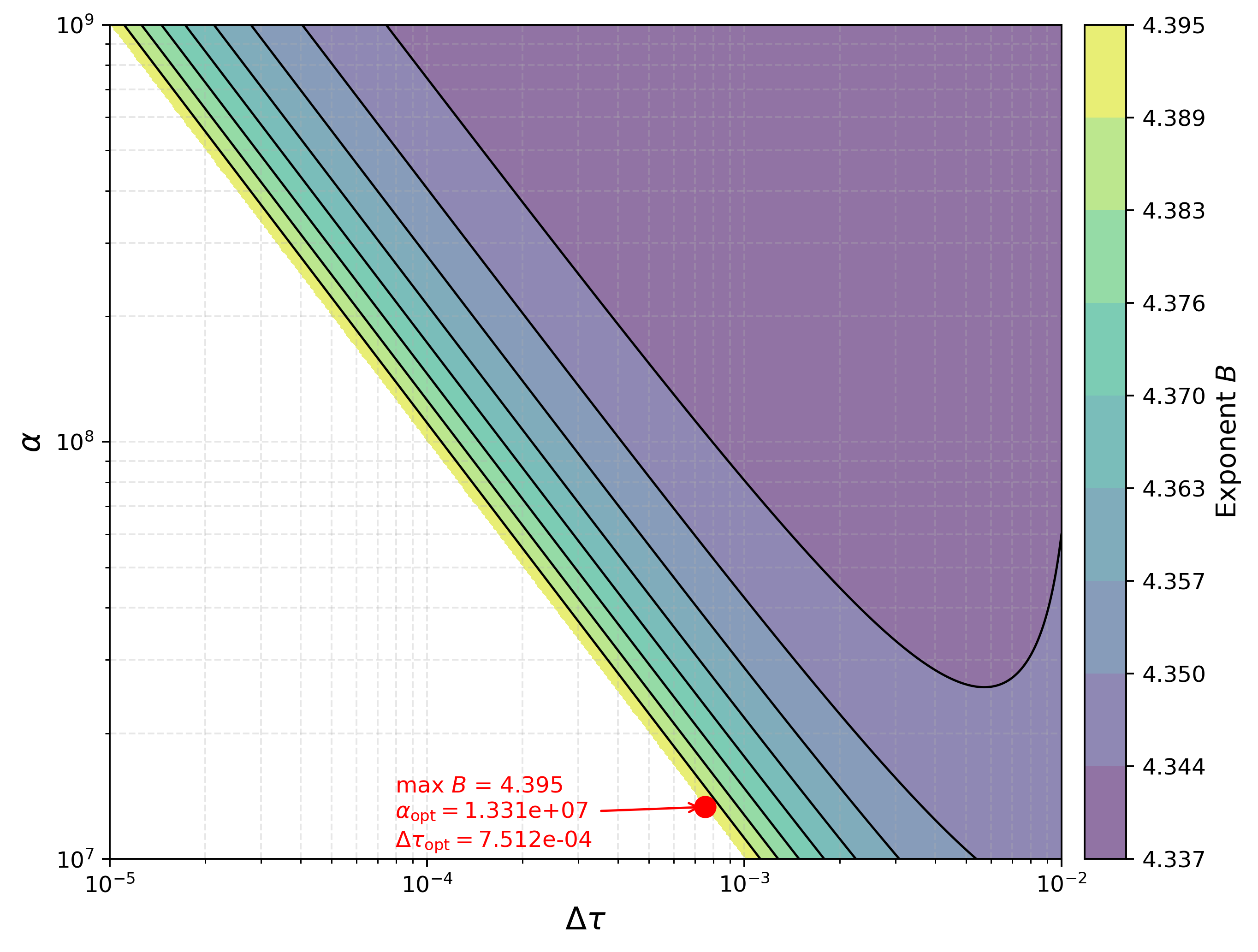}
    \caption{\small Maximum slope value as function of $\alpha$ and $\Delta\tau$ restricted such that $ 10^4 \leq \alpha \Delta \tau \leq 10^{8}$. Global maximum slope found to be $B = 4.395$ at $\Delta\tau = 7.5 \times 10^{-4}$, $\alpha = 1.33 \times 10^7$.}
    \label{gradsearch}
\end{figure}
\noindent
  Figure \ref{gradsearch} demonstrates the variation of the power-law exponent with the choices of $\alpha$ and $\Delta\tau$.  In the limit that $\Delta\tau \rightarrow 0$ and $\alpha \rightarrow \infty $ we find that $B \rightarrow 4.4$. For this subspace, we find the global maximum slope be $B = 4.398$ for $\Delta\tau = 7 \times 10^{-4}$, $\alpha = 1.33 \times 10^7$. Observably, the resulting slope is smaller
  than the value $B=4.75$ quoted in the main
  text, obtained in the limit \eqref{eq_laral}. We interpret the discrepancy as due to the
  fact that in this appendix we assume a power-law behaviour with constant $B$ for the entire growing part
  of the spectrum, however it is anticipated that the corresponding
   spectral index has a non-trivial
   dependence on the scale -- as seen
   for example in Fig \ref{fig_growthMF2}, right panel. It is intended that a more
   systematic analysis of this topic will 
 be left to future work.

\end{appendix}

{\small


\providecommand{\href}[2]{#2}\begingroup\raggedright\endgroup

}

\end{document}